\documentclass[12pt]{article}
\usepackage[margin=1in,footskip=0.25in]{geometry}

\usepackage{graphicx}
\usepackage{amssymb}
\usepackage{lscape}
\usepackage{caption}
\usepackage{subcaption}
\usepackage{xcolor}
\usepackage{enumitem}
\usepackage{textcomp}
\usepackage{mathtools}
\usepackage{comment}
\usepackage[normalem]{ulem}

\usepackage[hyphens]{url}
\usepackage[hidelinks]{hyperref}
\hypersetup{breaklinks=true}

\usepackage{array}
\usepackage{xcolor}
\usepackage{multirow}

\usepackage{tabularx}
\usepackage{adjustbox}
\usepackage{multicol}
\usepackage{longtable}

\usepackage{authblk}

\newcolumntype{C}[1]{>{\centering\arraybackslash\hspace{0pt} }p{#1} }
\newcolumntype{R}[1]{>{\raggedleft\arraybackslash\hspace{0pt} }p{#1} }
\newcolumntype{L}[1]{>{\raggedright\arraybackslash\hspace{0pt} }p{#1} }
\newcolumntype{P}[1]{>{\hspace{0pt} }p{#1}}

\usepackage[authoryear, round]{natbib}

\begin{document}
\bibliographystyle {plainnat}

\title{The Impact of an Employee’s Psychological Contract Breach on Compliance with Information Security Policies: Intrinsic and Extrinsic Motivation}

\author[1]{\small Daeun Lee\thanks{Daeun.Lee@warwick.ac.uk}}
\author[2]{\small Harjinder Singh Lallie\thanks{HL@warwick.ac.uk}}
\author[3]{\small Nadine Michaelides\thanks{Nadine.Michaelides@warwick.ac.uk}}

\affil[2]{Cyber Security Centre, WMG, University of Warwick, Coventry, CV4 7AL}

\date{}
\maketitle

\title{}

\begin{abstract}
Despite the rapid rise in social engineering attacks, not all employees are as compliant with information security policies (ISPs) to the extent that organisations expect them to be. ISP non-compliance is caused by a variety of psychological motivation. This study investigates the effect of psychological contract breach (PCB) of employees on ISP compliance intention (ICI) by dividing them into intrinsic and extrinsic motivation using the theory of planned behaviour (TPB) and the general deterrence theory (GDT). Data analysis from UK employees (\textit{n=206}) showed that the higher the PCB, the lower the ICI. The study also found that PCBs significantly reduced intrinsic motivation (attitude and perceived fairness) for ICI, whereas PCBs did not moderate the relationship between extrinsic motivation (sanction severity and sanctions certainty) and ICI. As a result, this study successfully addresses the risks of PCBs in the field of IS security and proposes effective solutions for employees with high PCBs.

\end{abstract}


\noindent \textbf{Competing interests.} The authors declare that we have no competing financial or non-financial interests that are directly or indirectly related to the work submitted for publication.

\section{Introduction}\label{Introduction} 

Organisational information security breaches can largely be explained by human error and omission \citep{RN102}. In other words, if employees deliberately or unintentionally fail to keep the information safe, it is insufficient to take technical countermeasures for the protection of the information. Accordingly, various psychological factors motivating employees’ failure to comply with ISP compliance have been raised in the cyber security literature. Among them, the Psychological Contract (PC) was presented as one of the significant human factors provoking employees’ cybersecurity behaviours \citep{RN178,RN104}. The PC is a set of beliefs about reciprocal obligations between an employee and an employer \citep{RN90}. According to the existing research, psychological contract breaches (PCBs) provoke poor organisational citizenship behaviours \citep{RN258} and even poor work performance \citep{RN94}. These results imply that employees’ PCBs are likely to reduce their ISP compliance intentions. However, empirical studies concerning the direct correlation between PCB and ISP compliance intentions have not been sufficiently conducted to date.

This research aims to evaluate the impact of PCB, a new potential psychological factor, on deficient ISP compliance intentions. The research also measures the impact of PCB on intrinsic and extrinsic motivation towards ISP compliance intentions in order to multifacetedly examine the risks of PCB. Consequently, the study can address the important role of PCB in IS security and provide a set of suggestions for employees having PCBs.

The rest of this paper is structured as follows. Section \ref{Background} aims to analyse the existing literature on PCB and ISP compliance intention to develop research hypotheses. Section \ref{Research Methodology} presents data analysis and results based on the research framework. The discussion in Section \ref{Discussion} proceeds to interpret and analyse the results to answer the research questions. Finally, Section \ref{Conclusions} describes the conclusions, recommendations, and limitations of this study.

\section{Background}\label{Background} 

\subsection{Psychological Contract}
Psychological contract has emerged as one of the most crucial factors in workforce management. Unlike the documented contract, the psychological contract is the unwritten contract and refers to an individual’s beliefs about mutual obligations between an employee and an organisation \citep{rousseau1989psychological}. When an employee perceives that the organisation is obliged to reciprocity for his or her contributions, the psychological contract is created. The contract has been constituted by paid-for-promises (e.g. high salary, promotion, long-term job security, or career development) made in exchange for some either implied or stated consideration such as hard work, accepting training, or transfers. Thus, psychological contracts are viewed as unwritten promises not as expectations. This leads employees to feel disappointed when psychological contracts are breached \citep{robinson1994violating}.

The consequences of psychological contract breaches have been found to negatively impact perceived obligations towards an employer, citizenship behaviour, commitment, satisfaction, intentions to remain and even work performance \citep{robinson1996trust, robinson1994violating, robinson1994changing}. For example, employees who experienced PCB do not tend to contribute to their organisation since they have no expectation of future benefit, which is the organisation’s obligation. Moreover, extreme cases of psychological contract breach could result in retaliation, sabotage, identity theft, and aggressive behaviour \citep{morrison1997employees}. Recent empirical studies have found PCB to negatively impact organisational behaviour \citep{RN978,mai2016examining}, job satisfaction, commitment and intention to leave \citep{RN259}, user resistance for the information system implementation \citep{RN261}, trust in organisation \citep{RN1142}, and productive work behaviour \citep{RN264}. PCB could lead to cybercrime conducted as a result of insider threat brought about by the PCB. However, this has not been thoroughly investigated.

\subsection{The Relationship Between Psychological Contracts and Intention to comply with Information Security Policies}

ISP (Information Security Policy) refers to any document that covers security programs, system controls and user behaviour within an organisation to realise security objectives \citep{landoll2017information}. ISP can be categorised into four levels: organisational-level policies, security program-level policies, user-level policies, and system and control-level policies. Among these, the present study focuses on user-level policies in order to identify an employee’s psychological factors that influence their behaviour and intentions. According to \textit{ISO (International Standards Organisation) 27001/2}, user-level policies consist of eight elements; security responsibility agreement, acceptable use of assets, security awareness program, removable media disposal procedures, document control plan, mobile device security policy, telework security policy, and disciplinary process \citep{landoll2017information}.

As cybercrime increases and becomes more severe and sophisticated, organisations put greater effort into information security risk management by implementing security measures and policies. Nonetheless, not only is the establishment of ISP within the organisation required, but employees must actively comply with ISP, playing a key role in substantially protecting cyber threats. Especially these days when social engineering is prevalent, the importance of encouraging employees to conform to ISP is increasingly emphasised \citep{flores2016shaping}. Therefore, it is expected that not only the information systems but also the users are obliged to adhere to the ISP statements.

However, if employees do not understand the importance of ISP compliance and are not willing to comply with it, all the technical measures and strategies that organisations have put in place will be in vain \citep{herath2009protection}. Hence, human factors affecting ISP compliance intentions are needed to be understood to encourage their motivation.

The PCB has been proposed as one of the most important factors influencing employees to perform security behaviours and to comply with security procedures. \cite{RN104} stated that employees are psychologically pressured to act in accordance with the expectations of the organisation by voluntarily limiting and maintaining their behaviours within the range of accepted practices. Therefore, if employees feel that the company breached their psychological contract, they could feel exasperated and compelled to get even with the company. In addition, \cite{abraham2011information} proposed PCB as one of the most influential factors associated with psychological ownership, organisational commitment, trust, as well as procedural justice.

While the necessity of investigating the impact of PCB in IS security has been increased, relevant empirical studies have not been sufficiently conducted. To the best of our knowledge there has been only one relevant empirical study: \cite{RN111} examined the mediating role of PCF (Psychological Contract Fulfillment) between perceived costs and ISP compliance intentions. The study conducted quantitative research seperated into supervisor and supervisee groups. As a result, it was found that PCF mitigates the negative impact of perceived costs on ISP compliance intentions only in the supervisor group. However, in this study, the perceived cost had no significant influence on ISP compliance intentions in both supervisor and supervisee groups. Accordingly, the study presents the hypothesis below.

\emph{\textbf{H1}: High Psychological Contract Breach has a strong negative effect on ISP compliance intentions.}

\subsection{Motivational Factors for ISP compliance intentions}
Extensive research has been done to examine human factors which influence employee compliance with ISP. Many behavioural theories (e.g. TPB (Theory of Planned Behaviour), GDT (General Deterrence Theory), PMT (Protection Motivation Theory), SCT (Social Cognitive Theory)) in IS literature have addressed motivators affecting ISP compliance. According to systematic literature reviews on behavioural theories, the most frequently used theory in IS security, was TPB followed by GDT \citep{alias2019information, lebek2013employees, lebek2014information}.

The TPB suggests that an individual’s behavioural intentions are determined by self-direction along with efforts to perform a target behaviour, or by motivation in terms of conscious plan and decision \citep{conner2020theory}. The TPB is mainly composed of attitudes, self-efficacy, and subjective norms. Attitudes are an individual’s overall assessments of a target behaviour, and self-efficacy is an individual’s expectation of how well they can control the target behaviour. Additionally, subjective norm is a function of normative beliefs which are an individual’s perceptions of the preferences of those around him who believe he should engage in targeted behaviour \citep{conner2020theory}. These three components are the most important psychological factors in motivating and predicting ISP compliance behaviours and intentions \citep{lebek2014information, RN1401}.

On the other hand, the GDT explains that a psychological process is made by deterring criminal behaviour only when people perceive that legal sanctions are clear, expeditious and harsh \citep{RN1398}. The GDT primarily consists of sanction severity and sanction certainty; sanction severity refers to an individual’s perception that penalties for non-compliance are severe and sanction certainty indicates an individual’s perception that risk of delinquent behaviour to be detected is high \citep{RN1398, safa2019deterrence}. 

\subsection{Intrinsic and Extrinsic Motivation}
People are motivated both internally and externally to take certain actions. Organisations typically seek to establish external measures such as sanctions and penalties for deviant cybersecurity behaviours, rather than increasing employees’ internal motivations. Extrinsic Motivation is defined as decision-making based on external factors such as a reward, surveillance, and punishment \citep{benabou2003intrinsic} as opposed to Intrinsic Motivation, which is an inherent desire to undertake the work even without specific rewards \citep{benabou2003intrinsic, makki2017influence}.

However, intrinsic and extrinsic motivation sometimes conflict with each other. According to \cite{benabou2003intrinsic}, some researchers insist that extrinsic motivations such as sanctions and rewards are often counterproductive since they often impede intrinsic motivation. This is because extrinsic motivations have limited effect on current employee engagement and reduces motivation to perform the same task later without compensation. Therefore, many social psychology studies emphasise the necessity to increase employee self-esteem rather than increase extrinsic motivation \citep{benabou2003intrinsic}. Accordingly, the study compares the effects of PCB on intrinsic and extrinsic motivation for ISP compliance intentions to identify how to motivate people who have experienced PCB to adhere to ISP.

\subsubsection{Intrinsic Motivations}
A psychological contract breach is known to induce negative emotional responses, which in turn reduces intrinsic motivation at work \citep{de2011m, morrison1997employees}. Conversely, it has been shown that psychological contract fulfilment increases motivation towards organisational commitment \citep{berman2003psychological}. Therefore, the present study suggests PCB negatively influences intrinsic motivation towards ISP compliance intentions.

The study adopted attitudes and self-efficacy of TPB as intrinsic motivators of ISP compliance intentions. This is because attitude has been studied as the most significant intrinsic motivator \citep{bulgurcu2011information}, and intrinsic motivation consists of autonomy and competence, which are aligned with self-efficacy \citep{alzahrani2018information}.

Additionally, employee psychological contract violations have been found to provoke negative organisational attitudes (e.g. job satisfaction, effective commitment, turnover intentions) \citep{RN1403, RN1405}. On the other hand, the correlation between perceived contract violation and low job satisfaction was found to be weaker as the work-related self-efficacy increased \citep{RN1406}. Therefore, it is necessary to study the mitigating role of self-efficacy on the negative effects of PCB. 

Employees who have experienced psychological contract breach may think that following the ISP is important but unfair, which may unwittingly lead to inadequate cybersecurity. Perceived Fairness can be defined as an individual’s perception of the fairness of an organisation’s ISP requirements, that exists within the internal context of ISP compliance \citep{bulgurcu2011information}. Perceived fairness has been found positively affect attitudes towards ISP compliance \citep{bulgurcu2009roles, bulgurcu2011information}. 
In terms of the relationship between perceived fairness and PCB, some research has found that employees’ beliefs of unfairness in the organisation’s regulations and treatments can be directly linked to psychological contract violation \citep {harrington2015drives, morrison1997employees}. Moreover, psychological contract fulfilment has been found to raise employees’ perception of performance appraisal fairness \citep {harrington2015drives}. It was also found that higher perceived fairness mitigated the negative influence of PCB on those with violated feelings \citep{RN261}. Hence, the study additionally measures an employee’s perceived fairness towards ISP compliance as an intrinsic motivational factor. Accordingly, the study proposes the following hypotheses:

\emph{\textbf{H2}: Higher intrinsic motivation (Attitudes, Self-efficacy, and Perceived Fairness) has a stronger positive effect on ISP compliance intentions.}

\emph{\textbf{H3a}: There is a negative effect of Psychological Contract Breach on Attitudes towards ISP compliance intentions.}

\emph{\textbf{H3b} There is a negative effect of Psychological Contract Breach on Self-efficacy towards ISP compliance intentions.}

\emph{\textbf{H3c}: H3c: There is a negative effect of Psychological Contract Breach on Perceived Fairness towards ISP compliance intentions.}

\subsubsection{Extrinsic Motivations}
Employees are sometimes compelled to follow organisational policies, even if they are unwilling to do so, to avoid disadvantages such as penalties and reputational damage. The study suggested subjective norm of TPB as well as sanction severity and sanction certainty of GDT as extrinsic motivators to influence employee compliance with ISP. Although some researchers have regarded subjective norms as somewhat voluntary behaviours, it has been considered as an extrinsic motivator in IS studies since intrinsic motivations are based on the employee’s desire to perform the task for himself or herself \citep{herath2009encouraging}. Therefore, subjective norm, sanction severity, and sanction certainty can be classified as extrinsic motivation factors in this study.

However, those extrinsic factors motivating employees to adhere to the policies can conflict with intrinsic motivation. According to a systematic literature review on IS behaviour theories, those extrinsic motivators of GDT – sanction severity and sanction certainty – have been found not to significantly influence IS deviant behaviours compared to TPB \citep{RN1401, safa2019deterrence}. This implies that the intrinsic motivators including PCB can either reduce the positive correlation between extrinsic motivation and ISP compliance intentions or reverse the direction of the positive correlation in a negative way. For instance, higher extrinsic motivation has a less positive effect or has no distinct effect on ISP compliance intentions when an employee’s PCB is high. Conversely, when an employee’s PCB is low, higher extrinsic motivation has a higher positive effect on ISP compliance intentions. Therefore, the study proposes the following hypotheses:

\emph{\textbf{H4}: High extrinsic motivation (Subjective norms, Sanction severity, and Sanction certainty) has a stronger positive effect on and ISP compliance intentions.}

\emph{\textbf{H5a}: Psychological contract breach moderates the relationship between Subjective Norms and ISP compliance intentions.}

\emph{\textbf{H5b}: Psychological contract breach moderates the relationship between Sanction Severity and ISP compliance intentions.}

\emph{\textbf{H5c}: Psychological contract breach moderates the relationship between Sanction Certainty and ISP compliance intentions.}

Lastly, since PCB can be classified as intrinsic motivation, it is assumed that PCB has a greater negative effect on people having intrinsic motivation than those having extrinsic motivation. Therefore, people who follow ISPs due to the external factors may not be relatively affected by the PCB since external factors are not changed by PCB. However, those who intrinsically seek to follow ISPs may be greatly affected by the PCB. Accordingly, the study suggests the following hypothesis:

\emph{\textbf{H6}: The effect of PCB on Intrinsic Motivation is stronger than the moderating effect of PCB between Extrinsic Motivation and ISP compliance intention.}

Consequently, the study combines two behaviour theories, TPB and GDT, classifying into intrinsic and extrinsic motivation based on the main research question, the negative impact of PCB on ISP compliance intentions. Accordingly, the study can differ the impact of PCB on intrinsic motivation to extrinsic motivation towards ISP compliance intentions. The proposed theoretical framework is presented in \textbf{Figure 1}.

\begin{figure}[!ht]
\begin{center}
    \includegraphics[width=1.0\linewidth]{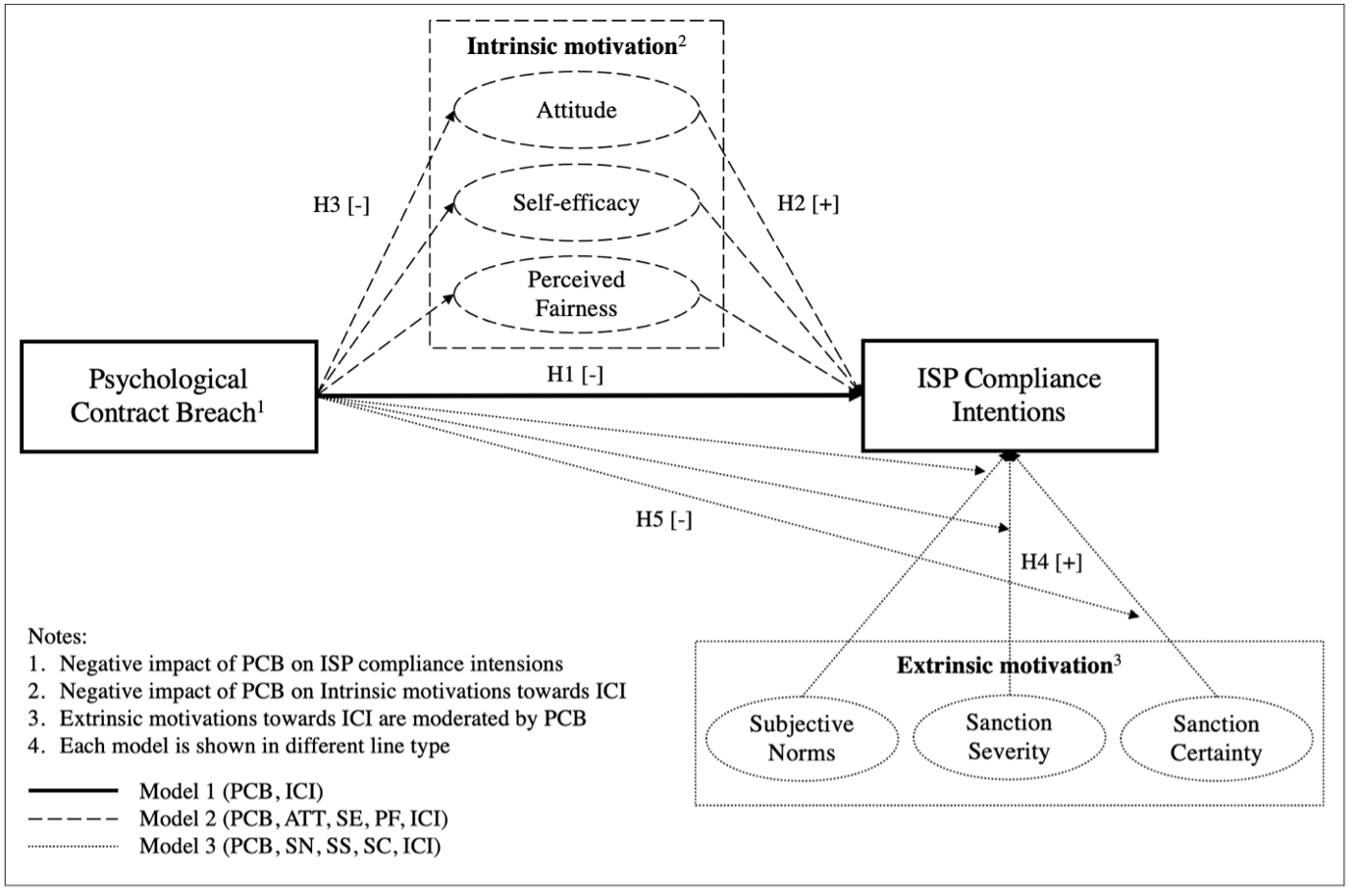}
    \label{fig:figure1}
    \caption{Proposed theoretical framework of the study}
    \end{center}
    \end{figure}

\subsection{Research Contribution}

In an era where most cyber-attack strategies target human weaknesses, it has become imperative for organisations to understand which human factors impact their employees’ security behaviour and foster their willingness to abide by the security regulations. However, enhancing employee intention requires more than providing a security awareness program. In order to properly comply with ISP, firstly, employees should be able to understand and practically apply the given information. Secondly, they should have attitudes and intentions to willingly comply with the policies \citep{bada2019cyber}. However, the attitudes and intentions to comply with ISP are accompanied by multifaceted psychological factors; employees’ evaluation of their capabilities to obey ISP (self-efficacy), disadvantages when not following the compliance (sanctions), and employees’ perceived expectations of coworkers (subjective norms) \citep{topa2015identifying}. Many behavioural theories have been researched in the field of IS security to date, grouping the relevant psychological factors. Among the various theories, this study will focus on social factors of TPB and GDT, dividing them into intrinsic and extrinsic motivational factors for ISP compliance intentions.

Meanwhile, although much research has found that many psychological factors affect ISP compliance behaviour and intentions, there are still potential factors that have not yet been properly studied. Likewise, no research has yet focused on the direct relation between PCB and ISP compliance intentions, although some theoretical studies \cite{RN178, RN104}  have implied the important role of PCB against complying with security policies. Conversely, one research study explored the mediating role of PCF (Psychological Contract Fulfilment) between perceived costs of Rational Choice Theory and ISP compliance intention. As a result of the study, the impact of PCF was influential in the supervisor group, but not prominent in the supervisee group \citep{RN111}. However, because they conducted the research with only limited factors, the influence of the PC in the non-administrator group was not thoroughly examined.

Accordingly, the research examines the research question: 
“How does an employee's Psychological Contract Breach affect
Information Security Policies Compliance Intentions?”

\section{Methodology}\label{Research Methodology}

\subsection{Data collection}
We used an online survey and recruited an FTSE 250 UK industrial goods and services company as a partner company for the survey. A single specific company was selected because it was important to ensure that participants were members of a company which had an appropriate ISP and that employees are aware of the ISP. Therefore, rather than distributing the survey to any employees, we decided to partner with a large corporation that provides a dedicated ISP. The survey was distributed only to employees working in the UK to facilitate communication and scheduling. A total sample size of 1,000 employees was selected through simple random sampling from a population of 3,021 employees of the partner company in the UK. As a result, 265 survey responses were received and only 208 responses were fully completed. Accordingly, the survey response rate was roughly 26.5\% and the survey completion rate was over 78.4\%. As a result of simply screening the data, there was two invalid responses within the 208 completed survey responses. Therefore, 206 completed responses remained valid for data analysis.

\subsection{Measures}
The questionnaire for this study was developed and combined, adopting reliable existing studies to collect quantitative data. The questionnaire is divided into the first part for the personal characteristics and the second part for the factors for substantial analysis. In part 1, a questionnaire for an employee’s demographic characteristics has been asked which are primarily identified as control variables in relevant empirical studies. Therefore, the following five variables have been included in the questionnaire. On the other hand, part 2 presents the substantial constructs of this study, consisting of 8 factors - Psychological Contract Breach (PCB), Attitudes (ATT), Self-efficacy (SE), Perceived Fairness (PF), Subjective norms (SN), Sanction Severity (SS), Sanction Certainty (SC), ISP Compliance Intention (ICI) - and 35 indicators. The full questionnaire is shown in \textbf{Appendix A}.


\subsection{Analysis and Results}
The data analysis was conducted, divided into 1) descriptive statistics for identifying personal characteristics, 2) measurement model analysis for construct validity and reliability, 3) structural model analysis for hypothesis testing, and 4) bivariate analysis for investigating the correlation between variables. The study employed IBM SPSS for descriptive statistics and SmartPLS 3.0 for confirmatory factor analysis (CFA). 

\subsubsection{Descriptive Statistics}
The personal characteristics collected in the Part 1 of the survey are shown in \textbf{Table 1}. The age group over the age of 19 is almost evenly distributed in all groups except for the oldest age group. Similarly, responses were received almost evenly from female and male respondents. By position, there were about twice as many non-managers as managers. Additionally, more than 40\% of respondents have worked for this organisation for one to five years and rates between about 9\% to 21\% have been shown in other tenure groups. Lastly, employee types have been divided into temporary and permanent type, with approximately 90\% of respondents were regular workers.

The normality test results of Part 2 are shown in \textbf{Table 7} in \textbf{Appendix B}. The mean value ranged from 1.42 (PCB8) to 2.23 (PCB4) for PCB and from 3.35 (SS2) to 4.83 (ICI1) for others. These statistics indicate that most respondents had moderately positive responses for the constructs of the study. The skewness value ranged from -2.862 (ATT2) to 2.354(PCB8), excluding ATT1 and ICI 1-3. Similarly, the kurtosis value ranged from -0.817 (PCB4) to 9.21(ATT2), except for ATT1 and ICI 1-4. ATT1 and ICI 1-4 failed the normality test since ATT1 and ICI 1-3 had absolute skewness values greater than or equal to 3.0, and ATT1 and ICI 1-4 had absolute kurtosis values greater than or equal to 10.0 \citep{brown2015confirmatory}. Therefore, a linear regression model, which is a non-parametric method that does not require normally distributed data, was additionally used in this study for variables that failed a normality test \citep{fathian2014analysis}.

Additionally, the descriptive statistics, including the mean, minimum and maximum values of PCB and ICI according to personal characteristics, are described in \textbf{Table 2}. Firstly, younger groups tend to have higher PCB. Additionally, the older group was more likely to comply with ISP overall, while the 20-29-year age group (4.78) had almost as high ICI as the 40-59-year age group (4.76). Secondly, managers (4.78) are more willing to comply with ISP than non-managers (4.73) although they have higher PCB. By tenure, the group with the shortest tenure had the lowest PCB (1.23) and ICI (4.71). Lastly, non-regular workers (1.45) had a much lower PCB level than regular workers (1.81), and their intention to comply with ISP (4.80) was higher than that of regular workers (4.74). Comparatively, there was no significant difference by gender in samples.

\begin{table}
\centering
\caption{Personal characteristics of the survey}
\footnotesize
\begin{tabular}{|l|l|l|l|l|}
\hline
Personal characteristics & Value & Frequency & Percent & Cumulative Percent \\ \hline
\multirow{6}{*}{Age} & Under 20 & 0 & 0 & 0 \\ \cline{2-5} 
 & 20-29 & 30 & 14.6 & 14.6 \\ \cline{2-5} 
 & 30-39 & 51 & 24.8 & 39.3 \\ \cline{2-5} 
 & 40-49 & 51 & 24.8 & 64.1 \\ \cline{2-5} 
 & 50-59 & 58 & 28.2 & 92.2 \\ \cline{2-5} 
 & 60 and above & 16 & 7.8 & 100 \\ \hline
\multirow{2}{*}{Gender} & Female & 99 & 48.1 & 48.1 \\ \cline{2-5} 
 & Male & 107 & 51.9 & 100 \\ \hline
\multirow{2}{*}{Job position} & Manager & 67 & 32.5 & 32.5 \\ \cline{2-5} 
 & Non-manager & 139 & 67.5 & 100 \\ \hline
\multirow{5}{*}{Tenure} & less than 1 year & 18 & 8.7 & 8.7 \\ \cline{2-5} 
 & 1-5 years & 86 & 41.7 & 50.5 \\ \cline{2-5} 
 & 6-10 years & 30 & 14.6 & 65 \\ \cline{2-5} 
 & 10-15 years & 28 & 13.6 & 78.6 \\ \cline{2-5} 
 & more than 15 years & 44 & 21.4 & 100 \\ \hline
\multirow{2}{*}{Employment type} & temporary & 21 & 10.2 & 10.2 \\ \cline{2-5} 
 & permanent & 185 & 89.8 & 100 \\ \hline
\end{tabular}%

\end{table}

\begin{table}
\centering
\footnotesize
\caption{Descriptive statistics for PCB and ICI according to the personal characteristics}
\begin{tabular}{|l|l|lll|lll|}
\hline
\multirow{2}{*}{Personal characteristics} & \multirow{2}{*}{Value} & \multicolumn{3}{l|}{PCB}                                     & \multicolumn{3}{l|}{ICI}                                     \\ \cline{3-8} 
                                          &                        & \multicolumn{1}{l|}{Mean} & \multicolumn{1}{l|}{Min.} & Max. & \multicolumn{1}{l|}{Mean} & \multicolumn{1}{l|}{Min.} & Max. \\ \hline
\multirow{6}{*}{Age}                      & Under 20               & \multicolumn{1}{l|}{N/A}  & \multicolumn{1}{l|}{N/A}  & N/A  & \multicolumn{1}{l|}{N/A}  & \multicolumn{1}{l|}{N/A}  & N/A  \\ \cline{2-8} 
                                          & 20-29                  & \multicolumn{1}{l|}{1.96} & \multicolumn{1}{l|}{1.00} & 4.33 & \multicolumn{1}{l|}{4.78} & \multicolumn{1}{l|}{3.25} & 5.00 \\ \cline{2-8} 
                                          & 30-39                  & \multicolumn{1}{l|}{1.89} & \multicolumn{1}{l|}{1.00} & 4.67 & \multicolumn{1}{l|}{4.60} & \multicolumn{1}{l|}{2.00} & 5.00 \\ \cline{2-8} 
                                          & 40-49                  & \multicolumn{1}{l|}{1.76} & \multicolumn{1}{l|}{1.00} & 4.78 & \multicolumn{1}{l|}{4.76} & \multicolumn{1}{l|}{2.00} & 5.00 \\ \cline{2-8} 
                                          & 50-59                  & \multicolumn{1}{l|}{1.69} & \multicolumn{1}{l|}{1.00} & 4.22 & \multicolumn{1}{l|}{4.76} & \multicolumn{1}{l|}{1.00} & 5.00 \\ \cline{2-8} 
                                          & 60 and above           & \multicolumn{1}{l|}{1.50} & \multicolumn{1}{l|}{1.00} & 3.44 & \multicolumn{1}{l|}{4.91} & \multicolumn{1}{l|}{4.50} & 5.00 \\ \hline
\multirow{2}{*}{Gender}                   & Female                 & \multicolumn{1}{l|}{1.75} & \multicolumn{1}{l|}{1.00} & 4.33 & \multicolumn{1}{l|}{4.74} & \multicolumn{1}{l|}{1.00} & 5.00 \\ \cline{2-8} 
                                          & Male                   & \multicolumn{1}{l|}{1.80} & \multicolumn{1}{l|}{1.00} & 4.78 & \multicolumn{1}{l|}{4.74} & \multicolumn{1}{l|}{2.00} & 5.00 \\ \hline
\multirow{2}{*}{Job position}             & Manager                & \multicolumn{1}{l|}{1.86} & \multicolumn{1}{l|}{1.00} & 4.78 & \multicolumn{1}{l|}{4.78} & \multicolumn{1}{l|}{2.00} & 5.00 \\ \cline{2-8} 
                                          & Non-manager            & \multicolumn{1}{l|}{1.74} & \multicolumn{1}{l|}{1.00} & 4.67 & \multicolumn{1}{l|}{4.73} & \multicolumn{1}{l|}{1.00} & 5.00 \\ \hline
\multirow{5}{*}{Tenure}                   & less than 1   year     & \multicolumn{1}{l|}{1.23} & \multicolumn{1}{l|}{1.00} & 2.00 & \multicolumn{1}{l|}{4.71} & \multicolumn{1}{l|}{3.00} & 5.00 \\ \cline{2-8} 
                                          & 1-5 years              & \multicolumn{1}{l|}{1.85} & \multicolumn{1}{l|}{1.00} & 4.67 & \multicolumn{1}{l|}{4.75} & \multicolumn{1}{l|}{3.00} & 5.00 \\ \cline{2-8} 
                                          & 6-10 years             & \multicolumn{1}{l|}{1.95} & \multicolumn{1}{l|}{1.00} & 4.78 & \multicolumn{1}{l|}{4.80} & \multicolumn{1}{l|}{2.00} & 5.00 \\ \cline{2-8} 
                                          & 10-15 years            & \multicolumn{1}{l|}{1.89} & \multicolumn{1}{l|}{1.00} & 3.89 & \multicolumn{1}{l|}{4.85} & \multicolumn{1}{l|}{4.00} & 5.00 \\ \cline{2-8} 
                                          & more than 15   years   & \multicolumn{1}{l|}{1.68} & \multicolumn{1}{l|}{1.00} & 3.56 & \multicolumn{1}{l|}{4.63} & \multicolumn{1}{l|}{1.00} & 5.00 \\ \hline
\multirow{2}{*}{Employment type}          & temporary              & \multicolumn{1}{l|}{1.45} & \multicolumn{1}{l|}{1.00} & 3.33 & \multicolumn{1}{l|}{4.80} & \multicolumn{1}{l|}{3.25} & 5.00 \\ \cline{2-8} 
                                          & permanent              & \multicolumn{1}{l|}{1.81} & \multicolumn{1}{l|}{1.00} & 4.78 & \multicolumn{1}{l|}{4.74} & \multicolumn{1}{l|}{1.00} & 5.00 \\ \hline
\end{tabular}%

\end{table}

\subsubsection{Inferential Statistics}
To verify construct validity and reliability, Confirmatory Factor Analysis (CFA) was performed in this study. The higher the factor loading size (0.8), the better the condition \citep{wiktorowicz2016exploratory}. However, since the factor loading of 0.4 has been also considered as significant \citep{factorloading}, the construct validity and reliability of all items in this study are significantly good or moderate, see \textbf{Table 8} in \textbf{Appendix C}. In \textbf{Table 9} of \textbf{Appendix C}, the constructs were additionally verified through multiple measurement models including Cronbach's Alpha, rho\_A, CR, and AVE.

Subsequently, the study analysed motivational process for ISP compliance intention with various constructs, by dividing into three structural models. The first model consists of PCB and ICI to investigate the direct correlation between them. In the second model, intrinsic motivation factors including ATT, SE, PF were added to the constructs in the first model. The third model was to determine the moderating effect of PCB on the relationship between extrinsic motivators such as SN, SS, SC, and ICI. The structural models were examined with T test, path coefficient, and P values to multifaceted investigate the relationships between factors. 

\textbf{Table 3} shows the result of the structural analysis for the hypotheses of this study. The direct correlation between PCB and ICI (\textbf{H1}) was verified to be strongly significant with p values of 0.002. Moreover, with the -0.195 value of path coefficient, the weak negative effect of PCB on ICI was shown. On the other hand, the relationship between intrinsic motivation and ICI (\textbf{H2}, \textbf{H3}) was assessed to be partially significant because only ATT among the three intrinsic motivators was significantly related to ICI. However, ATT-ICI and PCB-ATT relationships were found to have p values of 0.008 and 0.000 respectively. Additionally, indirect relationship of PCB-ATT-ICI had p values of 0.028. On the other hand, the impact of PCB was found to be the most significant on PF through t test and path coefficient along with p values. However, since PF-ICI relationship was not supported, the impact of PCB on PF towards ICI was not proved through a structural analysis. Therefore, while SE and PF were not found to be significant, the impact of PCB on ATT towards ISP compliance intentions were shown to be very strong. Lastly, among the three extrinsic motivators (\textbf{H4}, \textbf{H5}), the impacts of both SN and SC on ICI were very strong with p values of 0.000 and 0.001 respectively, while there was no relevance between SS and ICI. Comparatively, the moderating effect of PCB was not significant on SN, SS, as well as SC. \textbf{Figure 2} demonstrates the results of statistical analysis based on the theoretical framework of this study.


\begin{figure}[!b]
\begin{center}
    \includegraphics[width=1.0\linewidth]{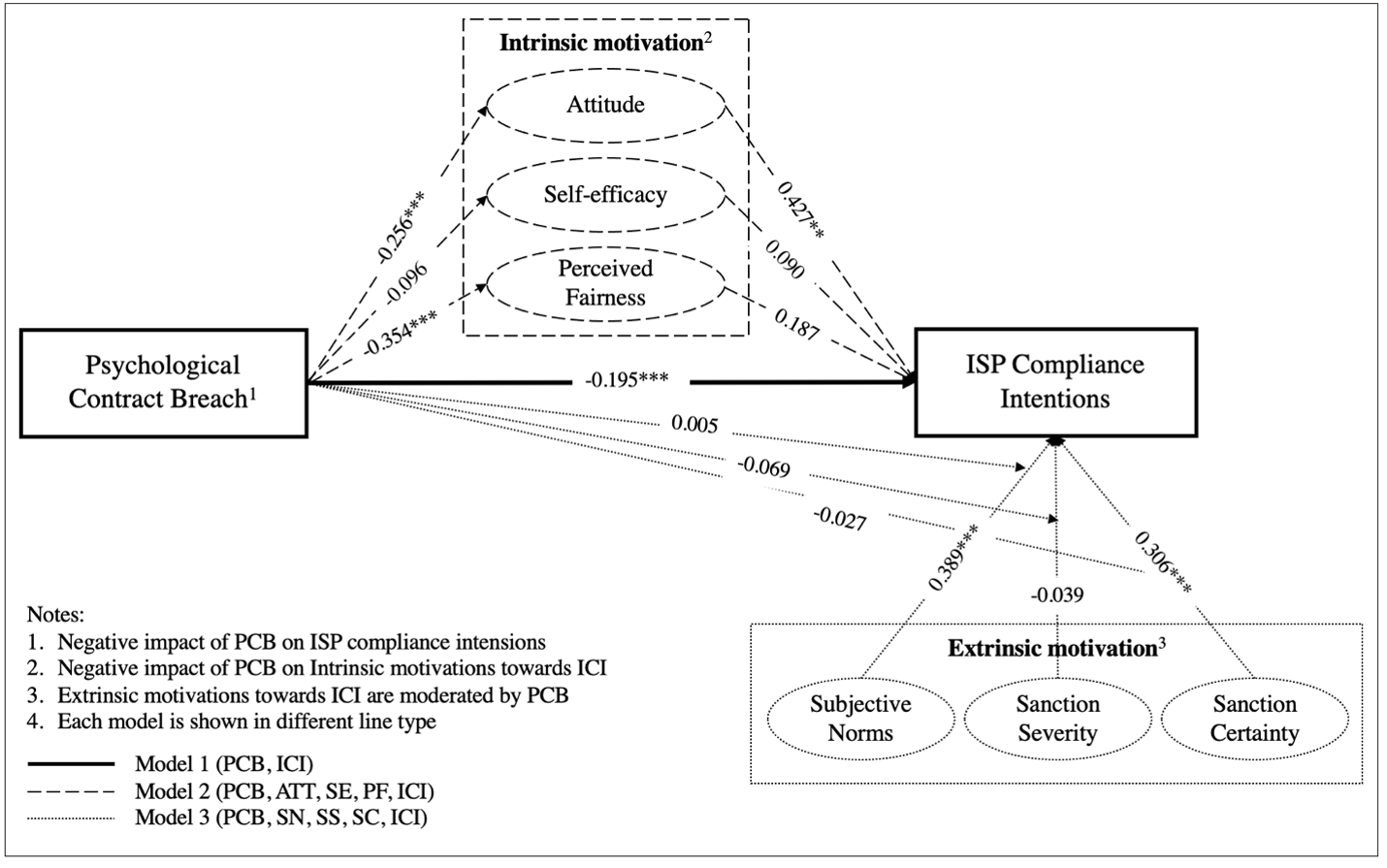}
    \caption{Structural statistics for theoretical framework of the study}
    \label{fig:figure2}
    \end{center}
    \end{figure}

\subsubsection{bivariate analysis}

\begin{table}[!t]
\centering
\footnotesize
\caption{Pearson correlation coefficient analysis between PCB, ATT, SE, and PF and ICI}
\begin{tabular}{lllllll}
\hline
\multicolumn{1}{|l|}{}                     & \multicolumn{1}{l|}{}                    & \multicolumn{1}{l|}{PCB}     & \multicolumn{1}{l|}{ATT}     & \multicolumn{1}{l|}{SE}     & \multicolumn{1}{l|}{PF}      & \multicolumn{1}{l|}{ICI}    \\ \hline
\multicolumn{1}{|l|}{\multirow{2}{*}{PCB}} & \multicolumn{1}{l|}{Pearson Correlation} & \multicolumn{1}{l|}{1}       & \multicolumn{1}{l|}{-.219**} & \multicolumn{1}{l|}{-0.078} & \multicolumn{1}{l|}{-.331**} & \multicolumn{1}{l|}{-.158*} \\ \cline{2-7} 
\multicolumn{1}{|l|}{}                     & \multicolumn{1}{l|}{Sig. (2-tailed)}     & \multicolumn{1}{l|}{0.002}   & \multicolumn{1}{l|}{0.268}   & \multicolumn{1}{l|}{0}      & \multicolumn{1}{l|}{0.023}   & \multicolumn{1}{l|}{}       \\ \hline
\multicolumn{1}{|l|}{\multirow{2}{*}{ATT}} & \multicolumn{1}{l|}{Pearson Correlation} & \multicolumn{1}{l|}{-.219**} & \multicolumn{1}{l|}{1}       & \multicolumn{1}{l|}{.300**} & \multicolumn{1}{l|}{.496**}  & \multicolumn{1}{l|}{.520**} \\ \cline{2-7} 
\multicolumn{1}{|l|}{}                     & \multicolumn{1}{l|}{Sig. (2-tailed)}     & \multicolumn{1}{l|}{0.002}   & \multicolumn{1}{l|}{}        & \multicolumn{1}{l|}{0}      & \multicolumn{1}{l|}{0}       & \multicolumn{1}{l|}{0}      \\ \hline
\multicolumn{1}{|l|}{\multirow{2}{*}{SE}}  & \multicolumn{1}{l|}{Pearson Correlation} & \multicolumn{1}{l|}{-0.078}  & \multicolumn{1}{l|}{.300**}  & \multicolumn{1}{l|}{1}      & \multicolumn{1}{l|}{.193**}  & \multicolumn{1}{l|}{.230**} \\ \cline{2-7} 
\multicolumn{1}{|l|}{}                     & \multicolumn{1}{l|}{Sig. (2-tailed)}     & \multicolumn{1}{l|}{0.268}   & \multicolumn{1}{l|}{0}       & \multicolumn{1}{l|}{}       & \multicolumn{1}{l|}{0.005}   & \multicolumn{1}{l|}{0.001}  \\ \hline
\multicolumn{1}{|l|}{\multirow{2}{*}{PF}}  & \multicolumn{1}{l|}{Pearson Correlation} & \multicolumn{1}{l|}{-.331**} & \multicolumn{1}{l|}{.496**}  & \multicolumn{1}{l|}{.193**} & \multicolumn{1}{l|}{1}       & \multicolumn{1}{l|}{.407**} \\ \cline{2-7} 
\multicolumn{1}{|l|}{}                     & \multicolumn{1}{l|}{Sig. (2-tailed)}     & \multicolumn{1}{l|}{0}       & \multicolumn{1}{l|}{0}       & \multicolumn{1}{l|}{0.005}  & \multicolumn{1}{l|}{}        & \multicolumn{1}{l|}{0}      \\ \hline
\multicolumn{1}{|l|}{\multirow{2}{*}{ICI}} & \multicolumn{1}{l|}{Pearson Correlation} & \multicolumn{1}{l|}{-.158*}  & \multicolumn{1}{l|}{.520**}  & \multicolumn{1}{l|}{.230**} & \multicolumn{1}{l|}{.407**}  & \multicolumn{1}{l|}{1}      \\ \cline{2-7} 
\multicolumn{1}{|l|}{}                     & \multicolumn{1}{l|}{Sig. (2-tailed)}     & \multicolumn{1}{l|}{0.023}   & \multicolumn{1}{l|}{0}       & \multicolumn{1}{l|}{0.001}  & \multicolumn{1}{l|}{0}       & \multicolumn{1}{l|}{}       \\ \hline
\multicolumn{7}{l}{\small** Significant at the 0.01 level (2-tailed).} \\
\multicolumn{7}{l}{\small* Significant at the 0.05 level (2-tailed).}                                                                                                                                                                               
\end{tabular}%
\end{table}

\begin{table}
\centering
\footnotesize
\caption{Pearson correlation coefficient analysis between SN, SS, and SC and ICI}
\begin{tabular}{lllllll}
\hline
\multicolumn{1}{|l|}{}                     & \multicolumn{1}{l|}{}                    & \multicolumn{1}{l|}{SN}     & \multicolumn{1}{l|}{SS}     & \multicolumn{1}{l|}{SC}     & \multicolumn{1}{l|}{ICI}    & \multicolumn{1}{l|}{ICI}    \\ \hline
\multicolumn{1}{|l|}{\multirow{2}{*}{SN}}  & \multicolumn{1}{l|}{Pearson Correlation} & \multicolumn{1}{l|}{1}      & \multicolumn{1}{l|}{.254**} & \multicolumn{1}{l|}{.346**} & \multicolumn{1}{l|}{.433**} & \multicolumn{1}{l|}{-.158*} \\ \cline{2-7} 
\multicolumn{1}{|l|}{}                     & \multicolumn{1}{l|}{Sig. (2-tailed)}     & \multicolumn{1}{l|}{}       & \multicolumn{1}{l|}{0}      & \multicolumn{1}{l|}{0}      & \multicolumn{1}{l|}{0}      & \multicolumn{1}{l|}{}       \\ \hline
\multicolumn{1}{|l|}{\multirow{2}{*}{SS}}  & \multicolumn{1}{l|}{Pearson Correlation} & \multicolumn{1}{l|}{.254**} & \multicolumn{1}{l|}{1}      & \multicolumn{1}{l|}{.530**} & \multicolumn{1}{l|}{.203**} & \multicolumn{1}{l|}{.520**} \\ \cline{2-7} 
\multicolumn{1}{|l|}{}                     & \multicolumn{1}{l|}{Sig. (2-tailed)}     & \multicolumn{1}{l|}{0}      & \multicolumn{1}{l|}{}       & \multicolumn{1}{l|}{0}      & \multicolumn{1}{l|}{0.003}  & \multicolumn{1}{l|}{0}      \\ \hline
\multicolumn{1}{|l|}{\multirow{2}{*}{SC}}  & \multicolumn{1}{l|}{Pearson Correlation} & \multicolumn{1}{l|}{.346**} & \multicolumn{1}{l|}{.530**} & \multicolumn{1}{l|}{1}      & \multicolumn{1}{l|}{.411**} & \multicolumn{1}{l|}{.230**} \\ \cline{2-7} 
\multicolumn{1}{|l|}{}                     & \multicolumn{1}{l|}{Sig. (2-tailed)}     & \multicolumn{1}{l|}{0}      & \multicolumn{1}{l|}{0}      & \multicolumn{1}{l|}{}       & \multicolumn{1}{l|}{0}      & \multicolumn{1}{l|}{0.001}  \\ \hline
\multicolumn{1}{|l|}{\multirow{2}{*}{ICI}} & \multicolumn{1}{l|}{Pearson Correlation} & \multicolumn{1}{l|}{.433**} & \multicolumn{1}{l|}{.203**} & \multicolumn{1}{l|}{.411**} & \multicolumn{1}{l|}{1}      & \multicolumn{1}{l|}{.407**} \\ \cline{2-7} 
\multicolumn{1}{|l|}{}                     & \multicolumn{1}{l|}{Sig. (2-tailed)}     & \multicolumn{1}{l|}{0}      & \multicolumn{1}{l|}{0.003}  & \multicolumn{1}{l|}{0}      & \multicolumn{1}{l|}{}       & \multicolumn{1}{l|}{0}      \\ \hline
\multicolumn{7}{l}{\small** Significant at the 0.01 level (2-tailed).}                                                                                                                                                                           
\end{tabular}%
\end{table}

\textbf{Figure 3} shows scattered plots with simple linear regression analyses. The PCB has negative correlation with all intrinsic motivation (ATT, SE, PF) as well as ICI, supporting hypotheses 1, 3, and 5. On the other hand, hypotheses 2 and 4 were supported by the positive correlation between ICI and all constructs except PCB (ATT, SE, PF, SN, SS, SC).

\begin{figure}
\begin{center}
    \includegraphics[width=0.9\linewidth]{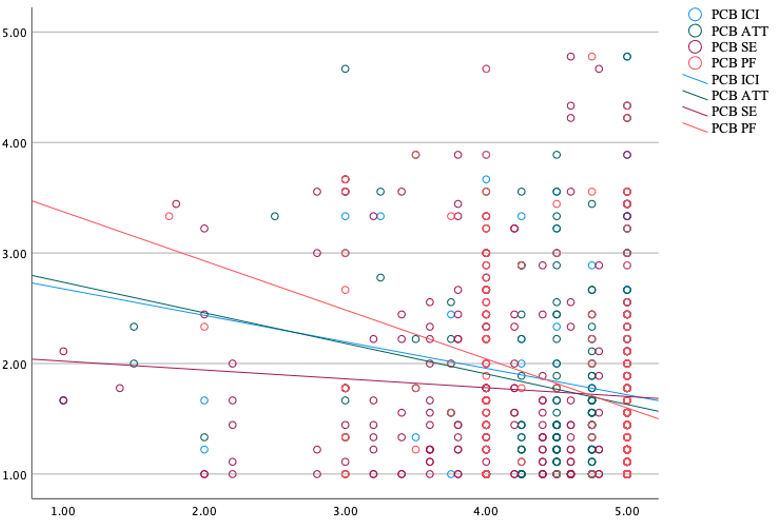}
    \includegraphics[width=0.9\linewidth]{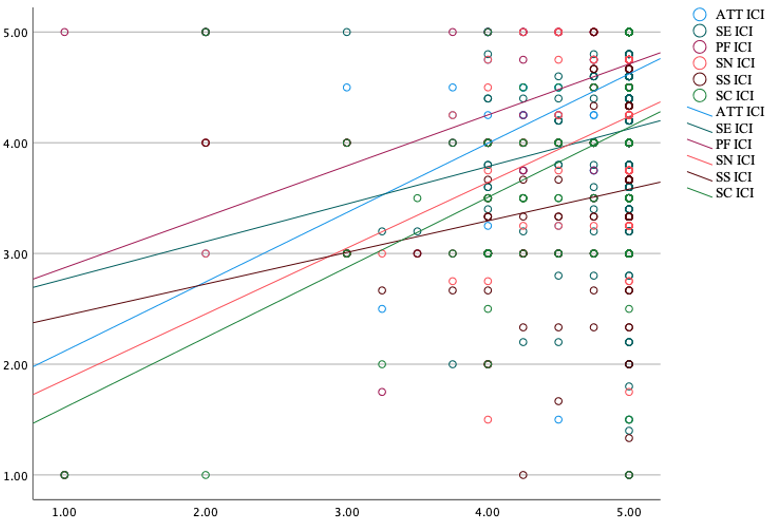}
    \caption{Linear regression analysis for the impact of PCB (top) and the predictors of ICI (bottom)}
    \label{fig:figure3}
    \end{center}
\end{figure}

\section{Discussion}\label{Discussion} 
As a result of the hypothesis test, it was found that the intention to comply with ISP was significantly affected by PCB, ATT, SN, and SC. Firstly, it has been shown that the higher the PCB of an employee, the more likely they are to be compliant with ISP. Of the three intrinsic motivators (ATT, SE, PF), only the ATT-ICI relationship was found to be significant while both SE and PF did not appear to affect ICI. In addition, PCB had a great negative effect on ATT and the indirect relationship of PCB-ATT-ICI was also found to be significant. Thus, PCB were found to have a negative impact on attitudes towards ISP compliance intentions. On the other hand, the PF-ICI relationship was too weak to support the hypothesis although PCB had a negative impact on PF for ISP. Accordingly, only the impact of PCB on ATT towards ICI was supported in the second model.

Among the three extrinsic motivators (SN, SC, SS), SN and SC showed a positive relation with ICI as expected by the existing theories. In contrast, the effect of SS on ICI was not significant. Additionally, the moderating role of PCB between the three factors and ICI was not significant at all, suggesting that PCB do not moderate the strongly positive SN-ICI and SC-ICI relationships. Subsequently, \textbf{H6} was significantly supported. Among intrinsic motivators, PCB negatively influenced ATT, which had a correlated effect on ICI to a significant extent. On the other hand, while SN and SC were found to affect ICI positively, they were not moderated by PCB. This result can be interpreted that PCB can reduce positive intrinsic motivation for ICI while PCB does not influence the extrinsic motivation for ICI. Therefore, the effect of PCB on intrinsic motivation is stronger than the moderating effect of psychological contract breach between extrinsic motivation and ISP compliance intention.

The Pearson correlation coefficient explained that all relationships in the theoretical framework of the study are correlated, except for PCB-SE relationship. Additionally, contrary to the structural analysis results, the PF-ICI and SS-ICI relationships were shown to have a significant positive correlation. Furthermore, as a result of simple linear regression analysis, PCB showed a negative correlation with intrinsic motivation and ICI, whereas all motivation factors except PCB have a positive correlation with ICI.

To sum up the results, it was confirmed that the negative correlation and causal relationship between PCB and ICI were significant, verifying hypothesis 1. These results can contribute to expanding existing research on the negative effects of PCB in organisations. Second, the study aimed to investigate how psychological factors such as intrinsic and extrinsic motivators for ICI could be negatively affected by PCB. As a result, ATT for ICI was significantly negatively affected by PCB, suggesting that PCB could decrease positive attitudes towards ICI. Lastly, it was shown that PCB did not moderate the positive correlation between extrinsic motivation and ICI.

Based on the findings, the study can propose that increasing intrinsic motivation and establishing extrinsic factors for prevent employees with PCB from performing inadequate cybersecurity behaviour. In particular, organisations should pay attention to fulfil their employees’ psychological contracts and strive to improve their attitudes for ISP compliance. Additionally, to address the risk of psychological contract breaches, organisations can encourage employees’ extrinsic motivation by building a cybersecurity culture and establishing certain sanctions for ISP compliance breaches.

\section{Conclusions}\label{Conclusions} 

Most cyber threat actors today often leverage human factors, as known as social engineering attacks or people hacking, which makes employee ISP compliance much more important. Nevertheless, not all employees are willing to be ISP compliant as the organisation expects them to be. Although most employees claim that they do not have enough time to comply with all ISPs during work, ISP noncompliance is rather driven by a variety of psychological motivations. Psychological contract breach has emerged as a major issue in the business environment because it fosters negative employee beliefs against the organisation. Therefore, the study conducted an empirical study to investigate the effect of PCB on ICI. In this study, the psychological factors of the Theory of Planned Behaviour and General Deterrence Theory were additionally applied by classifying it as intrinsic and extrinsic motivation. Data analysis primarily revealed that high PCB significantly led to low ISP compliance intentions. As a result, it was found that PCB greatly reduced intrinsic motivation (attitudes and perceived fairness) for ICI but did not moderate the relationship between extrinsic motivation (subjective norms and sanction certainty) and ICI. Overall, this study showed that an employees’ PCB played a significant role in influencing ISP compliance intentions.

\subsection{Recommendations}
Based on the findings, the study can propose that increasing intrinsic motivation and establishing extrinsic factors prevent employees with PCB from performing inadequate cybersecurity behaviour. In particular, organisations should pay attention to fulfilling their employees’ psychological contracts and strive to improve their attitudes for ISP compliance. Additionally, to address the risk of psychological contract breaches, organisations can encourage employee extrinsic motivation by building a cybersecurity culture and establishing certain sanctions for ISP compliance breaches.

In addition to establishment of ISP, employee ISP compliance is essential to avoid threats of people hacking and social engineering. Therefore, reducing PCB is important not only for employee engagement and work performance but also for information security risk management. The most important ways to address the risks of employee PCB is to make promises clear from the beginning. Alternatively, PCB can be mitigated by open communication, trust in the supervisor, and specific obligations (e.g. job content, career development, organisational policies, leadership and social contacts, work-life balance, job security, rewards) \citep{van2020role}. Besides, it was found that the relationship between PCB and work performance was moderated in employees having high social interaction, perceived organisational support, and trust \citep{RN1466}.

Second, organisations should strive to increase positive attitudes and perceived fairness. In addition to fulfilling the employee psychological contract, a manager’s persuasive strategy can increase employee attitudes and intrinsic motivation more effectively than an assertive strategy \citep{chiu2018employees}. In addition, organisations should identify why employees perceive that the ISP compliance requirements are unfair. Third, the study also suggests that an organisation’s cybersecurity culture can mitigate an employee’s undesirable security behaviours, which can be caused by high PCB. Organisations must make significant investments in implementing transformational change to build a cybersecurity culture that goes beyond simply offering a SETA program \citep{alshaikh2020developing}. In addition, despite the PCB, employees are inevitably inclined to comply with the ISP to avoid their misbehaviours getting caught up. Thus, the final proposal of this study is to pay attention to employee behaviour and ISP compliance. Organisations can also establish security measures to monitor and alert employees for breaches of security compliance. 

\subsection{Limitations and Directions for Future Research}
This study conducted a cross-sectional survey that measured only partial and static phenomenon due to the time frame of the study \citep{bravo2019psychological}. Therefore, the path coefficient was analysed in order to examine the causal relationship between PCB and motivators as well as ICI. Nevertheless, the study was unable to identify whether PCB was created before other psychological factors. Accordingly, a longitudinal study is proposed to be employed in future research.

In addition, the average value of PCB collected from the partner company was very low (1.78), while the average ICI was very high (4.74). Therefore, this might have affected the significance of the impact of PCB on ICI. Out of the 206 valid responses, only 30 employees had PCBs of 3.0 or higher and 176 employees had PCBs less than 3.0. Therefore, the study was unable to divide the sample into breached and non-breached groups. Furthermore, only 3 employees had an ICI of less than 3, while 203 employees had an ICI of 3 or much higher. Such biased data could have affected the significance of relationships between factors. Thus, in future research, it is desirable to recruit multiple companies to diversify the range of PCB and ICI. The structural model analysis found that SE, PF, and SS were not significant for ICI. There were also no significant effects of SE for ICI with correlation efficient although SE has been long studied to have very strong correlation with ICI in IS studies \citep{lebek2014information,RN1401}. On the contrary, PF and SS showed a significant correlation with ICI through correlation coefficient analysis. This can lead to the conclusion that the relationship between PCB and the three factors were not fully investigated. Therefore, these relationships should be further investigated in future studies, especially in longitudinal studies. 


\bibliography{bibliography.bib}

\newpage


\begingroup

\newpage
\LARGE{Appendix A}
\\
\\
\small{
PART ONE: Personal Characteristics\\
1.	What is your age?\\
{[}  {]} under 20\\
{[}  {]} 20-29\\
{[}  {]} 30-39\\
{[}  {]} 40-49\\
{[}  {]} 50-59\\
{[}  {]} 60 and above\\
\\
2.	What is your gender?\\
{[}  {]} Female\\
{[}  {]} Male\\
\\
3.	What is your job position?\\
{[}  {]} Manager\\
{[}  {]} Non-manager\\
\\
4.	How long have you worked in this organisation?\\
{[}  {]} less than 1 year\\
{[}  {]} 1-5 years\\
{[}  {]} 6-10 years\\
{[}  {]} 10-15 years\\
{[}  {]} more than 15 years\\
\\
5.	What is your type of employment?\\
{[}  {]} Temporary\\
{[}  {]} Permanent\\
}
\\
\clearpage
\noindent
PART TWO: Motivational process for ISP compliance intention\\
\small{To what extent do you agree? \\ * ISP (Information Security Policy) prescribes employee's cybersecurity behaviour within an organisation (e.g. use of personal computers, access to the internal systems, opening emails and attachments, data leakage from social media, password management, and software downloads from the internet)}.


\begin{longtable}{|p{0.2\linewidth}|p{0.07\linewidth}|p{0.45\linewidth}|p{0.15\linewidth}|}
\caption{Research Questionnaire}
    \\
    \hline
    \textbf{Factor} & \textbf{Item} & \textbf{Item Description} & \textbf{Sources} \\\cline{2-4}
    \hline
    \multirow{9}{\linewidth}{Psychological contract breach} & PCB1 & Almost all the promises made by my employer during recruitment have been kept so far. & \multirow{9}{\linewidth}{\cite{RN978,RN90}} \\ \cline{2-3} 
    & PCB2 & I feel that my employer has come through in fulfilling the promises made to me when I was hired. &\\\cline{2-3}
    & PCB3 & So far my employer has done an excellent job of fulfilling its promises to me. &\\\cline{2-3}
    & PCB4 & I have not received everything promised to me in exchange for my contribution. &\\\cline{2-3}
    & PCB5 & My employer has broken many of its promises to me even though I’ve upheld my side of the deal. &\\\cline{2-3}
    & PCB6 & I feel a great deal of anger toward my organisation. &\\\cline{2-3}
    & PCB7 & I feel betrayed by my organisation. &\\\cline{2-3}
    & PCB7 & I feel that my organisation has violated the contract between us. &\\\cline{2-3}
    & PCB9 & I feel extremely frustrated by how I have been treated by my organisation. &\\
    \hline
    \multicolumn{3}{|p{0.4\linewidth}}{Intrinsic Motivation} & \\
    \hline
    \multirow{4}{\linewidth}{Attitudes} & ATT1 & Following the organisation’s ISP is a good idea. & \multirow{4}{\linewidth}{\cite{ifinedo2012understanding}} \\ \cline{2-3}
    & ATT2 & Following the organisation’s ISP is a necessity. & \\\cline{2-3}
    & ATT3 & Following the organisation’s ISP is beneficial. & \\\cline{2-3}
    & ATT4 & Following the organisation’s ISP is pleasant. & \\
    \hline
    \multirow{5}{\linewidth}{Self-efficacy} & SE1 & I would feel comfortable following most of the ISP on my own. & \multirow{5}{\linewidth}{\cite{herath2009protection,ifinedo2012understanding}} \\ \cline{2-3}
    & SE2 & If I wanted to, I could easily follow ISP by on my own. & \\\cline{2-3}
    & SE3 & I would be able to follow most of the ISP even if there was no one around to help me. & \\\cline{2-3}
    & SE4 & I believe that it is within my control to protect myself from information security violations. & \\
    & SE5 & I have the necessary skills to protect myself from information security violations. & \\
    \hline
    \multirow{4}{\linewidth}{Perceived Fairness} & PF1 & I believe the requirements of the ISP that I am required to comply with are unfair. & \multirow{4}{\linewidth}{\cite{bulgurcu2010quality}} \\ \cline{2-3}
    & PF2 & I believe the requirements of the ISP that I am required to comply with are unreasonable. & \\\cline{2-4}
    & PF3 & I believe the expectations of the organisation that I should comply with the ISP is unfair. &\multirow{2}{\linewidth}{Self-developed} \\\cline{2-3}
    & PF4 & I believe the expectations of the organisation that I should comply with the ISP is unreasonable. & \\
    \hline
    \multicolumn{3}{|p{0.4\linewidth}}{Extrinsic Motivation} & \\
    \hline
    \multirow{4}{\linewidth}{Subjective Norms} & SN1 & My boss thinks that I should follow the organisation’s ISP. & \multirow{4}{\linewidth}{\cite{ifinedo2012understanding}} \\ \cline{2-3}
    & SN2 & My colleagues think that I should follow the organisation’s ISP. & \\\cline{2-3}
    & SN3 & My organisation’s IT department pressures me to follow the organisation’s ISPs. & \\\cline{2-3}
    & SN4 & My subordinates think I should follow the organisation’s ISP. & \\
    \hline
    \multirow{3}{\linewidth}{Sanction Severity} & SS1 & The organisation disciplines employees who break information security rules. & \multirow{3}{\linewidth}{\cite{herath2009protection}} \\ \cline{2-3}
    & SS2 & My organisation terminates employees who repeatedly break security rules. & \\\cline{2-3}
    & SS3 & If I were caught violating organisation information security policies, I would be severely punished. & \\\cline{2-3}
    \hline
    \multirow{2}{\linewidth}{Sanction Certainty} & SC1 & Employee computer practices are properly monitored for policy violations. & \multirow{2}{\linewidth}{\cite{herath2009protection}} \\ \cline{2-3}
    & SC2 & If I violate organisation security policies, I would probably be caught. & \\\cline{2-4}
    \hline
    \multirow{4}{\linewidth}{ISP Compliance Intentions} & ICI1 & I intend to follow the organisation’s ISP. & \multirow{4}{\linewidth}{\cite{herath2009protection,RN550}} \\ \cline{2-3}
    & ICI2 & I am likely to follow the organisation’s ISP. & \\\cline{2-3}
    & ICI3 & It is possible that I will comply with ISP to protect the organisation’s information systems. & \\\cline{2-3}
    & ICI4 & I am certain that I will follow organisational ISP. & \\
    \hline
      \end{longtable}
\vspace{-0.4cm}
\hspace{-0.6cm}
Note - 1: Strongly disagree, 2: Somewhat disagree, 3: Neither agree nor disagree, 4: Somewhat agree, 5: Strongly agree
    \label{tab:Table 5.1}

\clearpage

\newpage\LARGE{Appendix B}

\small{For analysis, the values of PCB 1-3 were reverse coded as PCB represent a negative factor, whereas PCB 4-9 remained the same. Similar, the values of PF 1-4 were reverse coded as perceived fairness was a positive motivator. Therefore, a low value for the PCB indicator can be interpreted as positive, while a high value is associated with a positive factor for the other 25 indicators.}\\

\begin{table}[h]
\centering
\caption{Results showing normality test of factors}
\resizebox{\textwidth}{!}{%
\begin{tabular}{|l|l|l|l|ll|l|l|l|l|}
\hline
\multirow{2}{*}{Indicator} & \multirow{2}{*}{\begin{tabular}[c]{@{}l@{}}Range\\    Statistics\end{tabular}} & \multirow{2}{*}{Min.   Statistics} & \multirow{2}{*}{\begin{tabular}[c]{@{}l@{}}Max.\\    \\ Statistics\end{tabular}} & \multicolumn{2}{l|}{Mean}                    & \multirow{2}{*}{Std. Deviation Statistic} & \multirow{2}{*}{Variance Statistics} & \multirow{2}{*}{Skewness} & \multirow{2}{*}{Kurtosis} \\ \cline{5-6}
                           &                                                                                &                                    &                                                                                  & \multicolumn{1}{l|}{Statistics} & Std. Error &                                           &                                      &                           &                           \\ \hline
PCB1                       & 4                                                                              & 1                                  & 5                                                                                & \multicolumn{1}{l|}{1.92}       & 0.072      & 1.04                                      & 1.081                                & 1.218                     & 1.055                     \\ \hline
PCB2                       & 4                                                                              & 1                                  & 5                                                                                & \multicolumn{1}{l|}{1.96}       & 0.073      & 1.047                                     & 1.096                                & 1.067                     & 0.579                     \\ \hline
PCB3                       & 4                                                                              & 1                                  & 5                                                                                & \multicolumn{1}{l|}{2.02}       & 0.069      & 0.988                                     & 0.975                                & 0.851                     & 0.286                     \\ \hline
PCB4                       & 4                                                                              & 1                                  & 5                                                                                & \multicolumn{1}{l|}{2.23}       & 0.091      & 1.303                                     & 1.699                                & 0.677                     & -0.817                    \\ \hline
PCB5                       & 4                                                                              & 1                                  & 5                                                                                & \multicolumn{1}{l|}{1.75}       & 0.077      & 1.101                                     & 1.212                                & 1.302                     & 0.617                     \\ \hline
PCB6                       & 4                                                                              & 1                                  & 5                                                                                & \multicolumn{1}{l|}{1.52}       & 0.068      & 0.976                                     & 0.953                                & 1.805                     & 2.414                     \\ \hline
PCB7                       & 4                                                                              & 1                                  & 5                                                                                & \multicolumn{1}{l|}{1.52}       & 0.07       & 1.006                                     & 1.012                                & 1.95                      & 2.898                     \\ \hline
PCB8                       & 4                                                                              & 1                                  & 5                                                                                & \multicolumn{1}{l|}{1.42}       & 0.062      & 0.889                                     & 0.791                                & 2.354                     & 5.352                     \\ \hline
PCB9                       & 4                                                                              & 1                                  & 5                                                                                & \multicolumn{1}{l|}{1.67}       & 0.081      & 1.163                                     & 1.352                                & 1.562                     & 1.179                     \\ \hline
ATT1                       & 4                                                                              & 1                                  & 5                                                                                & \multicolumn{1}{l|}{4.71}       & 0.049      & 0.706                                     & 0.498                                & -3.145                    & 11.337                    \\ \hline
ATT2                       & 4                                                                              & 1                                  & 5                                                                                & \multicolumn{1}{l|}{4.68}       & 0.051      & 0.734                                     & 0.539                                & -2.862                    & 9.21                      \\ \hline
ATT3                       & 4                                                                              & 1                                  & 5                                                                                & \multicolumn{1}{l|}{4.62}       & 0.054      & 0.773                                     & 0.597                                & -2.365                    & 5.831                     \\ \hline
ATT4                       & 4                                                                              & 1                                  & 5                                                                                & \multicolumn{1}{l|}{3.83}       & 0.07       & 1.011                                     & 1.023                                & -0.414                    & -0.469                    \\ \hline
SE1                        & 4                                                                              & 1                                  & 5                                                                                & \multicolumn{1}{l|}{4.14}       & 0.064      & 0.913                                     & 0.834                                & -1.331                    & 1.945                     \\ \hline
SE2                        & 4                                                                              & 1                                  & 5                                                                                & \multicolumn{1}{l|}{3.93}       & 0.068      & 0.97                                      & 0.942                                & -1.06                     & 1.052                     \\ \hline
SE3                        & 4                                                                              & 1                                  & 5                                                                                & \multicolumn{1}{l|}{3.91}       & 0.072      & 1.034                                     & 1.07                                 & -1.096                    & 0.956                     \\ \hline
SE4                        & 4                                                                              & 1                                  & 5                                                                                & \multicolumn{1}{l|}{4.14}       & 0.067      & 0.955                                     & 0.912                                & -1.44                     & 2.321                     \\ \hline
SE5                        & 4                                                                              & 1                                  & 5                                                                                & \multicolumn{1}{l|}{4.07}       & 0.065      & 0.935                                     & 0.873                                & -1.222                    & 1.631                     \\ \hline
PF1                        & 4                                                                              & 1                                  & 5                                                                                & \multicolumn{1}{l|}{4.53}       & 0.051      & 0.737                                     & 0.543                                & -1.892                    & 4.577                     \\ \hline
PF2                        & 4                                                                              & 1                                  & 5                                                                                & \multicolumn{1}{l|}{4.55}       & 0.051      & 0.736                                     & 0.542                                & -1.727                    & 3.06                      \\ \hline
PF3                        & 3                                                                              & 2                                  & 5                                                                                & \multicolumn{1}{l|}{4.65}       & 0.044      & 0.629                                     & 0.396                                & -1.812                    & 3.011                     \\ \hline
PF4                        & 3                                                                              & 2                                  & 5                                                                                & \multicolumn{1}{l|}{4.65}       & 0.043      & 0.621                                     & 0.386                                & -1.801                    & 3.084                     \\ \hline
SN1                        & 4                                                                              & 1                                  & 5                                                                                & \multicolumn{1}{l|}{4.54}       & 0.055      & 0.788                                     & 0.62                                 & -1.641                    & 2.11                      \\ \hline
SN2                        & 4                                                                              & 1                                  & 5                                                                                & \multicolumn{1}{l|}{4.31}       & 0.064      & 0.922                                     & 0.849                                & -1.186                    & 0.726                     \\ \hline
SN3                        & 4                                                                              & 1                                  & 5                                                                                & \multicolumn{1}{l|}{3.54}       & 0.084      & 1.212                                     & 1.469                                & -0.493                    & -0.545                    \\ \hline
SN4                        & 4                                                                              & 1                                  & 5                                                                                & \multicolumn{1}{l|}{3.95}       & 0.071      & 1.023                                     & 1.046                                & -0.555                    & -0.319                    \\ \hline
SS1                        & 4                                                                              & 1                                  & 5                                                                                & \multicolumn{1}{l|}{3.44}       & 0.066      & 0.949                                     & 0.901                                & -0.125                    & 0.09                      \\ \hline
SS2                        & 4                                                                              & 1                                  & 5                                                                                & \multicolumn{1}{l|}{3.35}       & 0.061      & 0.875                                     & 0.765                                & 0.138                     & 0.781                     \\ \hline
SS3                        & 4                                                                              & 1                                  & 5                                                                                & \multicolumn{1}{l|}{3.73}       & 0.065      & 0.938                                     & 0.88                                 & -0.41                     & 0.121                     \\ \hline
SC1                        & 4                                                                              & 1                                  & 5                                                                                & \multicolumn{1}{l|}{3.87}       & 0.065      & 0.939                                     & 0.882                                & -0.531                    & 0.069                     \\ \hline
SC2                        & 4                                                                              & 1                                  & 5                                                                                & \multicolumn{1}{l|}{4.08}       & 0.065      & 0.936                                     & 0.876                                & -0.995                    & 0.884                     \\ \hline
ICI1                       & 4                                                                              & 1                                  & 5                                                                                & \multicolumn{1}{l|}{4.83}       & 0.039      & 0.56                                      & 0.314                                & -4.567                    & 24.429                    \\ \hline
ICI2                       & 4                                                                              & 1                                  & 5                                                                                & \multicolumn{1}{l|}{4.73}       & 0.049      & 0.708                                     & 0.501                                & -3.386                    & 12.741                    \\ \hline
ICI3                       & 4                                                                              & 1                                  & 5                                                                                & \multicolumn{1}{l|}{4.69}       & 0.053      & 0.765                                     & 0.586                                & -3.096                    & 10.39                     \\ \hline
ICI4                       & 4                                                                              & 1                                  & 5                                                                                & \multicolumn{1}{l|}{4.72}       & 0.042      & 0.598                                     & 0.357                                & -2.858                    & 10.697                    \\ \hline
\end{tabular}}
\end{table}

\newpage\LARGE{Appendix C}
\begin{table}[!ht]
\centering
\caption{Factor loadings and Cross-loading}
\begin{tabular}{|l|l|l|l|l|l|l|l|l|}
\hline
       & ATT                                   & ICI                                   & PCB                                   & PF                                    & SC                                    & SE                                    & SN                                    & SS                                    \\ \hline
ATT\_1 & {\color[HTML]{000000} \textbf{0.941}} & 0.517                                 & -0.223                                & 0.445                                 & 0.291                                 & 0.247                                 & 0.418                                 & 0.126                                 \\ \hline
ATT\_2 & {\color[HTML]{000000} \textbf{0.948}} & 0.504                                 & -0.24                                 & 0.495                                 & 0.29                                  & 0.253                                 & 0.417                                 & 0.124                                 \\ \hline
ATT\_3 & {\color[HTML]{000000} \textbf{0.93}}  & 0.527                                 & -0.246                                & 0.525                                 & 0.346                                 & 0.287                                 & 0.394                                 & 0.155                                 \\ \hline
ATT\_4 & {\color[HTML]{000000} \textbf{0.584}} & 0.308                                 & -0.172                                & 0.259                                 & 0.321                                 & 0.23                                  & 0.203                                 & 0.19                                  \\ \hline
ICI\_1 & 0.467                                 & {\color[HTML]{000000} \textbf{0.895}} & -0.141                                & 0.347                                 & 0.372                                 & 0.204                                 & 0.446                                 & 0.159                                 \\ \hline
ICI\_2 & 0.403                                 & {\color[HTML]{000000} \textbf{0.902}} & -0.154                                & 0.348                                 & 0.338                                 & 0.168                                 & 0.438                                 & 0.18                                  \\ \hline
ICI\_3 & 0.415                                 & {\color[HTML]{000000} \textbf{0.795}} & -0.099                                & 0.288                                 & 0.267                                 & 0.137                                 & 0.357                                 & 0.146                                 \\ \hline
ICI\_4 & 0.552                                 & {\color[HTML]{000000} \textbf{0.82}}  & -0.218                                & 0.432                                 & 0.476                                 & 0.312                                 & 0.406                                 & 0.26                                  \\ \hline
PCB\_1 & -0.312                                & -0.194                                & {\color[HTML]{000000} \textbf{0.871}} & -0.368                                & -0.376                                & -0.109                                & -0.339                                & -0.21                                 \\ \hline
PCB\_2 & -0.3                                  & -0.191                                & {\color[HTML]{000000} \textbf{0.889}} & -0.361                                & -0.361                                & -0.107                                & -0.33                                 & -0.195                                \\ \hline
PCB\_3 & -0.27                                 & -0.174                                & {\color[HTML]{000000} \textbf{0.876}} & -0.347                                & -0.334                                & -0.074                                & -0.287                                & -0.161                                \\ \hline
PCB\_4 & -0.081                                & -0.021                                & {\color[HTML]{000000} \textbf{0.619}} & -0.176                                & -0.179                                & -0.015                                & -0.113                                & -0.084                                \\ \hline
PCB\_5 & -0.214                                & -0.137                                & {\color[HTML]{000000} \textbf{0.824}} & -0.258                                & -0.286                                & -0.084                                & -0.286                                & -0.185                                \\ \hline
PCB\_6 & -0.091                                & -0.121                                & {\color[HTML]{000000} \textbf{0.724}} & -0.197                                & -0.255                                & -0.073                                & -0.172                                & -0.235                                \\ \hline
PCB\_7 & -0.141                                & -0.151                                & {\color[HTML]{000000} \textbf{0.79}}  & -0.247                                & -0.265                                & -0.061                                & -0.241                                & -0.257                                \\ \hline
PCB\_8 & -0.028                                & -0.076                                & {\color[HTML]{000000} \textbf{0.772}} & -0.197                                & -0.199                                & -0.035                                & -0.247                                & -0.237                                \\ \hline
PCB\_9 & -0.171                                & -0.172                                & {\color[HTML]{000000} \textbf{0.853}} & -0.263                                & -0.278                                & -0.074                                & -0.239                                & -0.264                                \\ \hline
PF\_1  & 0.461                                 & 0.376                                 & {\color[HTML]{000000} -0.367}         & {\color[HTML]{000000} \textbf{0.916}} & 0.344                                 & 0.185                                 & 0.416                                 & 0.139                                 \\ \hline
PF\_2  & 0.445                                 & 0.365                                 & -0.295                                & {\color[HTML]{000000} \textbf{0.908}} & 0.353                                 & 0.192                                 & 0.353                                 & 0.155                                 \\ \hline
PF\_3  & 0.512                                 & 0.415                                 & -0.319                                & {\color[HTML]{000000} \textbf{0.958}} & 0.33                                  & 0.182                                 & 0.412                                 & 0.099                                 \\ \hline
PF\_4  & 0.499                                 & 0.425                                 & -0.342                                & {\color[HTML]{000000} \textbf{0.967}} & 0.353                                 & 0.184                                 & 0.418                                 & 0.129                                 \\ \hline
SC\_1  & 0.265                                 & 0.356                                 & -0.338                                & 0.327                                 & {\color[HTML]{000000} \textbf{0.907}} & 0.219                                 & 0.335                                 & 0.474                                 \\ \hline
SC\_2  & 0.378                                 & 0.445                                 & -0.34                                 & 0.351                                 & {\color[HTML]{000000} \textbf{0.941}} & 0.194                                 & 0.405                                 & 0.52                                  \\ \hline
SE\_1  & 0.306                                 & 0.234                                 & -0.096                                & 0.18                                  & 0.176                                 & {\color[HTML]{000000} \textbf{0.868}} & 0.315                                 & 0.132                                 \\ \hline
SE\_2  & 0.243                                 & 0.216                                 & -0.026                                & 0.17                                  & 0.149                                 & {\color[HTML]{000000} \textbf{0.879}} & 0.236                                 & 0.1                                   \\ \hline
SE\_3  & 0.223                                 & 0.187                                 & -0.01                                 & 0.106                                 & 0.156                                 & {\color[HTML]{000000} \textbf{0.829}} & 0.245                                 & 0.092                                 \\ \hline
SE\_4  & 0.234                                 & 0.204                                 & -0.161                                & 0.186                                 & 0.236                                 & {\color[HTML]{000000} \textbf{0.853}} & 0.31                                  & 0.174                                 \\ \hline
SE\_5  & 0.244                                 & 0.231                                 & -0.09                                 & 0.191                                 & 0.221                                 & {\color[HTML]{000000} \textbf{0.872}} & 0.298                                 & 0.206                                 \\ \hline
SN\_1  & 0.431                                 & 0.513                                 & -0.274                                & 0.429                                 & 0.388                                 & 0.318                                 & {\color[HTML]{000000} \textbf{0.888}} & 0.193                                 \\ \hline
SN\_2  & 0.406                                 & 0.438                                 & -0.343                                & 0.424                                 & 0.416                                 & 0.322                                 & {\color[HTML]{000000} \textbf{0.932}} & 0.248                                 \\ \hline
SN\_3  & 0.128                                 & 0.207                                 & -0.145                                & 0.106                                 & 0.073                                 & 0.044                                 & {\color[HTML]{000000} \textbf{0.507}} & 0.146                                 \\ \hline
SN\_4  & 0.287                                 & 0.28                                  & -0.245                                & 0.287                                 & 0.296                                 & 0.282                                 & {\color[HTML]{000000} \textbf{0.784}} & 0.27                                  \\ \hline
SS\_1  & 0.125                                 & 0.182                                 & -0.224                                & 0.097                                 & 0.491                                 & 0.152                                 & 0.185                                 & {\color[HTML]{000000} \textbf{0.862}} \\ \hline
SS\_2  & 0.081                                 & 0.127                                 & -0.168                                & 0.025                                 & 0.4                                   & 0.132                                 & 0.131                                 & {\color[HTML]{000000} \textbf{0.834}} \\ \hline
SS\_3  & 0.187                                 & 0.237                                 & -0.226                                & 0.186                                 & 0.477                                 & 0.145                                 & 0.31                                  & {\color[HTML]{000000} \textbf{0.873}} \\ \hline
\end{tabular}
\end{table}
\clearpage
\newpage

\begin{table}[ht!]
\centering
\caption{Construct validity and reliability}
\begin{tabular}{|l|l|l|l|l|}
\hline
Construct & Cronbach's   Alpha & rho\_A & CR    & AVE   \\ \hline
ATT       & 0.877              & 0.918  & 0.92  & 0.748 \\ \hline
ICI       & 0.877              & 0.889  & 0.915 & 0.73  \\ \hline
PCB       & 0.935              & 0.97   & 0.943 & 0.65  \\ \hline
PF        & 0.954              & 0.957  & 0.967 & 0.879 \\ \hline
SC        & 0.831              & 0.861  & 0.921 & 0.854 \\ \hline
SE        & 0.913              & 0.921  & 0.934 & 0.74  \\ \hline
SN        & 0.801              & 0.902  & 0.868 & 0.632 \\ \hline
SS        & 0.826              & 0.878  & 0.892 & 0.734 \\ \hline 
\end{tabular}
\end{table}
\endgroup











\end{document}